# Ultra-High Electro-Optic Activity Demonstrated in a Silicon-Organic Hybrid (SOH) Modulator

CLEMENS KIENINGER,[1,2,*] YASAR KUTUVANTAVIDA,[1,2] DELWIN L. ELDER,[3]
STEFAN WOLF,[1] HEINER ZWICKEL,[1] MATTHIAS BLAICHER,[1] JUNED N. KEMAL,[1]
MATTHIAS LAUERMANN,[4] SEBASTIAN RANDEL,[1] WOLFGANG FREUDE,[1]
LARRY R. DALTON,[3] AND CHRISTIAN KOOS[1,2,*]

[1]*Karlsruhe Institute of Technology (KIT), Institute of Photonics and Quantum Electronics (IPQ), 76131 Karlsruhe, Germany*
[2]*Karlsruhe Institute of Technology (KIT), Institute of Microstructure Technology (IMT), 76344 Eggenstein-Leopoldshafen, Germany*
[3]*University of Washington, Department of Chemistry, Seattle, WA 98195, USA*
[4]*Now with: Infinera Corporation, Sunnyvale, CA 94089, USA*
*\*Corresponding authors: clemens.kieninger@kit.edu, christian.koos@kit.edu*

**Efficient electro-optic (EO) modulators crucially rely on advanced materials that exhibit strong electro-optic activity and that can be integrated into high-speed and efficient phase shifter structures. In this paper, we demonstrate ultra-high in-device EO figures of merit of up to $n^3 r_{33}$ = 2300 pm/V achieved in a silicon-organic hybrid (SOH) Mach-Zehnder Modulator (MZM) using the EO chromophore JRD1. This is the highest material-related in-device EO figure of merit hitherto achieved in a high-speed modulator at any operating wavelength. The π-voltage of the 1.5 mm-long device amounts to 210 mV, leading to a voltage-length product of $U_\pi L$ = 320 Vµm – the lowest value reported for MZM that are based on low-loss dielectric waveguides. The viability of the devices is demonstrated by generating high-quality on-off-keying (OOK) signals at 40 Gbit/s with $Q$ factors in excess of 8 at a drive voltage as low as 140 mV$_{pp}$. We expect that efficient high-speed EO modulators will not only have major impact in the field of optical communications, but will also open new avenues towards ultra-fast photonic-electronic signal processing.**

Electro-optic (EO) phase modulators are key for many applications such as optical communications [1], optical signal processing [2], or high-precision metrology [3]. Among the various concepts, devices based on the linear electro-optic effect (Pockels effect) are particularly powerful, providing pure phase modulation at high speed, thereby enabling the generation of high-quality optical data signals based on advanced modulation formats [4–6]. Pockels-effect modulators can be broadly subdivided in two categories: The first one comprises conventional devices, where optical waveguides are directly fabricated in crystalline or amorphous EO materials. This approach is used for, e.g., LiNbO$_3$ devices [7], all-polymer modulators [8], or devices based on III-V compound semiconductors [9,10], which often exploit a combination of the Pockels effect and other phase modulation mechanisms such as the quantum confined stark effect (QCSE) or the Franz Keldysh effect. More recently, a second category of devices has emerged, relying on hybrid integration approaches that combine Pockels-type EO materials with dielectric or plasmonic waveguide structures that do not feature any EO activity. This approach is often used to overcome the intrinsic lack of second-order nonlinearities in otherwise highly attractive material systems such as the silicon photonics platform. Examples of this category comprise the concepts of silicon-organic hybrid (SOH) [11–13] and plasmonic-organic hybrid (POH) [14–16] devices, that combine organic EO (OEO) materials with silicon photonic or plasmonic waveguide structures or other approaches that exploit hybrid combinations of lithium-niobate (LiNbO$_3$) thin films [17] or barium-titanate (BaTiO$_3$) layers [18] with silicon-on-insulator (SOI) waveguides.

In all of these cases, one of the most important design goals is to achieve high modulation efficiency, i.e., large phase shifts for low operating voltages and small device lengths, along with low optical insertion loss. For Mach-Zehnder modulators (MZM), the modulation efficiency can be quantified by the product $U_\pi L$ of the phase shifter length $L$ and the voltage $U_\pi$ that is required to achieve a phase shift of π between the optical signals at the output of the MZM arms [13]. For Pockels-effect modulators based on dielectric waveguides, the lowest demonstrated $U_\pi L$-product amounts to 0.5 Vmm, achieved in an SOH device that combines a silicon photonic slot waveguide with the OEO chromophore DLD164 [19]. For plasmonic waveguides, even lower values down to $U_\pi L = 0.05$ Vmm can be achieved [20]. However, while these devices feature large electro-optic bandwidths [21], they suffer from intrinsic propagation losses, which limit the range of practical device lengths to a few tens of micrometers [13,20]. This effect can be quantified by the product $aU_\pi L$ which combines the π-voltage $U_\pi$, the phase shifter length $L$, and the waveguide propagation loss coefficient $a$ measured in dB/mm. Previously demonstrated values of $aU_\pi L$ amount to 2.8 VdB for SOH devices [19], indicating that a phase shifter section with length $L$ and propagation loss $aL = 1$ dB requires an operating voltage of 2.8 V for a phase shift of π. For the previously mentioned POH devices [20], this number amounts to 25 VdB.

In principle, all these performance parameters can be improved by using Pockels-type materials that feature strong electro-optic activity, quantified by large EO figures of merit $n^3 r$, where $n$ is the refractive index and $r$ the EO coefficient. However, it is equally important that the materials offer good processing properties and allow easy application to efficient and high-speed phase shifter structures such as slot waveguides. The importance of fulfilling both requirements may be illustrated by the example of silicon-based $BaTiO_3$ modulators: These devices exhibit record-high in-device EO figures of merit of up to $n^3 r_{33} = 3200$ pm/V [18,22], but feature rather high voltage-length products in excess of $U_\pi L = 13.5$ Vmm. Moreover, $BaTiO_3$ modulators still suffer from low modulation speeds, with highest data rates below 1 Gbit/s [22]. The limited performance of these devices is directly linked to the complexity of the growth process of crystalline $BaTiO_3$, which can only be used in combination with rather simple phase shifter structures based on conventional strip waveguides.

In contrast to that, OEO materials are widely applicable and compatible to a wide variety of integration platforms and device structures, enabling high-speed modulators with record performance parameters [23–26] and data rates of up to 400 Gbit/s [5]. The materials can be easily spin-coated, micro-dispensed or printed in high-throughput processes. For devices based on OEO materials, the largest in-device EO figure of merit reported so far amounts to $n^3 r_{33} = 1220$ pm/V ($n^3 r_{33} = 2040$ pm/V), at a wavelength of $\lambda = 1.55$ μm ($\lambda = 1.25$ μm), achieved for an organic binary-chromophore compound in a POH modulator [27].

Here we demonstrate that even higher EO figures of merit of up to $n^3 r_{33} = 2\,300$ pm/V at $\lambda = 1.55$ μm can be achieved. The results were obtained in an SOH modulator that uses the organic EO material JRD1 [28] as a cladding. To the best of our knowledge, this is the highest in-device EO figure of merit ever achieved in a high-speed Pockels-type modulator at any operation wavelength. The devices feature $U_\pi L$-products down to 0.32 Vmm – the lowest value ever achieved for a modulator based on low-loss dielectric waveguides. We measure a $aU_\pi L$-product of 1.2 VdB, which is on par with the lowest values reported so far [10]. The viability of the device is demonstrated by generation of on-off-keying (OOK) signals at a data rate of 40 Gbit/s. For a peak-to-peak operation voltage as low as $U_{pp} = 140$ mV$_{pp}$, we obtain high-quality signals with measured $Q$ factors in excess of 8.

## SOH modulator concept and material design

The concept of an SOH phase shifter is depicted in Fig. 1(a). The device comprises a silicon (Si) slot waveguide formed by two 240 nm wide and 220 nm high Si rails that are separated by a narrow slot, which is filled by the OEO material. An electrical connection between the Si rails and the aluminum (Al) transmission lines is achieved by 70 nm high doped Si slabs and Al vias. Both optical and RF field are tightly confined in the slot region, leading to a strong modal overlap in the OEO material and hence to highly efficient phase modulation.

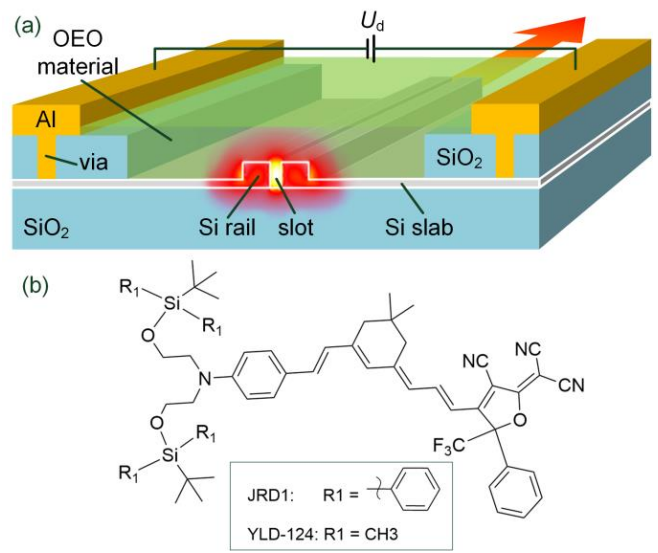

Fig. 1 Concept of a silicon-organic hybrid (SOH) phase shifter. (a) Device principle. The phase shifter comprises a Si slot waveguide formed by a pair of Si rails. The slot is filled with an organic electro-optic (OEO) material. The rails are electrically connected to an aluminum (Al) transmission line via doped Si slabs and Al vias, such that an externally applied voltage $U_d$ drops entirely across the narrow slot. At the same time, the optical mode is tightly confined to the OEO material in the slot, resulting in highly efficient phase modulation. (b) Molecular structure of the employed OEO material JRD1 and of the previously used [26] material YLD124. Compared to YLD124, JRD1 features bulky phenyl side groups denoted as R1, which decrease inter-molecular interactions and thus increase molecular mobility during electric-field poling. This is instrumental for achieving ultra-high in-device EO activity.

The waveguide structures are fabricated on standard silicon-on-insulator (SOI) wafers in a commercial silicon photonic foundry process using standard 248 nm deep-UV (DUV) lithography. As an OEO cladding, we use the neat chromophore material JRD1 [28], which is locally deposited on the slot waveguides in a dedicated post-processing step. The molecular structure of JRD1 is displayed in Fig. 1(b). The material is the result of theory-guided optimization of the previously used OEO chromophore YLD124 [26]. Both materials share common donor and acceptor groups that are separated by a π-conjugated bridge. Due to this common chromophore core, both materials show a very similar molecular hyperpolarizability [29]. However, JRD1 features bulky phenyl side groups denoted as R1 in Fig. 1(b), rather than methyl groups as used in YLD124. The bulkier side groups significantly increase the intermolecular distance of neighboring chromophores, which reduces the strength of detrimental dipole-dipole interactions that promote centrosymmetric pairing of chromophores [30]. As a result, JRD1 chromophores can be used in high number densities without a polymer host [30], while still maintaining high molecular mobility during electric-field poling and avoiding unwanted centrosymmetric pairing of chromophores. This leads to large EO figures of merit of $n^3 r_{33} = 3850$ pm/V in bulk samples of JRD1 [28] – more than a 2-fold increase compared to YLD124 [28]. Note that the glass transition temperatures of JRD1 ($T_g = 82$°C) and YLD124 ($T_g = 81$°C) are comparable [31]. For the JRD1-coated SOH modulator presented in this work, thermally induced relaxation of the chromophore orientation at room temperature does not impair the device performance significantly: A π-voltage increase of about 5% was recorded one week after the poling of the device. Thermal relaxation is most pronounced during the first few hundred hours [32,33], and a much weaker decay of $U_\pi$ is expected for larger storing times. Thermal stability at elevated temperatures without significant impairment of the EO activity can be achieved by modifying the material JRD1 with crosslinking agents which enable post-poling lattice hardening. This approach has led to similar classes of OEO materials with a $T_g$ of up to 250°C [34].

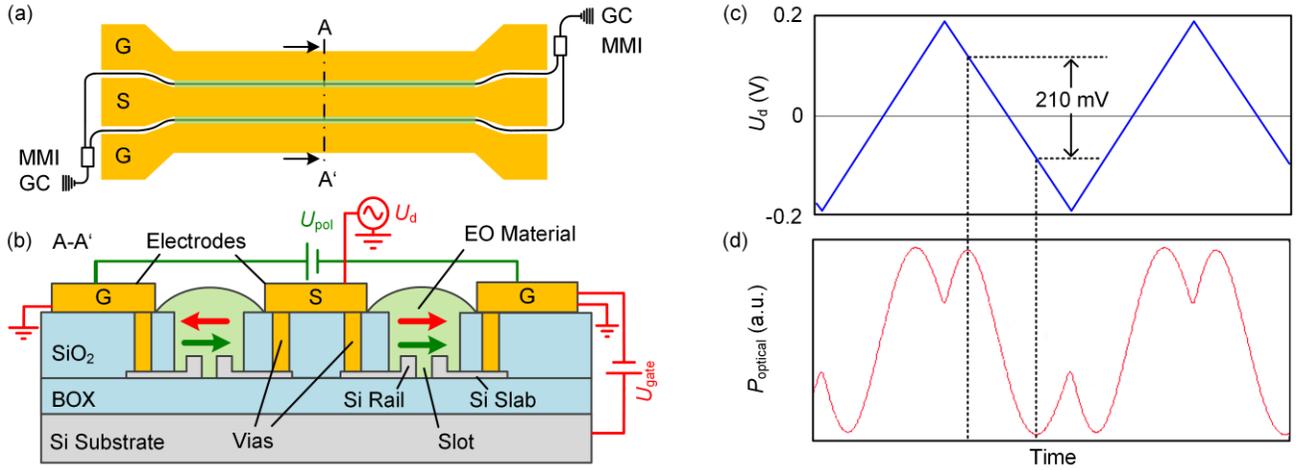

Fig. 2 Schematic of an SOH MZM and measurement of $U_\pi$. **(a)** Top view. Light is coupled to and from the device via grating couplers (GC). The coplanar transmission line is arranged in a ground-signal-ground (GSG) configuration. A pair of multi-mode interference (MMI) couplers is used to split and recombine the light of the two MZM arms. **(b)** Cross section of the two arms of the SOH MZM at the position defined by the dash-dotted line A-A' in (a). Each arm comprises an SOH phase shifter. A one-time poling process is required to achieve a macroscopic in-device $r_{33}$ after deposition of the EO chromophores from solution. To this end, a poling voltage $U_{pol}$ is applied across the floating ground electrodes at elevated temperature close to $T_g = 82°C$, thereby inducing an electric poling field (green arrows) in the slot regions. The dipoles of the EO material align along the poling field in acentric order. After cooling the device down to room temperature, $U_{pol}$ is removed and the acentric order of the chromophores is conserved. For modulation, a radio-frequency (RF) signal voltage $U_d$ induces electric fields in the slots (red arrows) that are antiparallel (parallel) to the aligned chromophores in the left (right) arm of the MZM, thereby realizing push-pull operation. The bandwidth of the modulator can be increased by decreasing the resistance of the silicon slab. This is achieved by applying a gate voltage $U_{gate}$ between the silicon substrate and the ground electrodes. **(c)** Low-speed triangular drive signal for the static $U_\pi$ measurement as a function of time. **(d)** Over-modulated optical signal as a function of time when the modulator is fed by the waveform depicted in (c). The MZM is biased in its quadrature point and the π-voltage $U_\pi$ can be directly read from the voltage increment needed to drive the device from minimum to maximum transmission. For the depicted measurements, we extract a π-voltage of 210 mV, corresponding to a π-voltage-length product of $U_\pi L = 0.32$ Vmm.

Note that the field enhancement of the optical mode in the silicon photonic slot waveguide leads to high intensities, which may damage the electro-optic chromophores. However, this effect is directly linked to the presence of oxygen in the material. It was previously reported that photo degradation can be prevented operating the device in an oxygen-free atmosphere [33,35]. This can be achieved by appropriate sealing techniques.

### Device preparation and characterization

For our experimental demonstration, we use Mach-Zehnder type SOH modulators rather than the phase modulator shown in Fig. 1(a). The concept of an SOH Mach-Zehnder modulator (MZM) is schematically depicted in Fig. 2(a). The coplanar transmission line is realized in a ground-signal-ground (GSG) configuration. Light is coupled to and from the chip via grating couplers (GC), and a multi-mode interference coupler (MMI) is used to split and combine the light of the two arms of the MZM. Fig. 2(b) shows a cross-section of the two MZM arms at the position indicated by the dash-dotted line A-A' in Fig. 2(a). Each MZM arm comprises an SOH phase shifter with a JRD1 cladding. The material is deposited in solution, thus the EO chromophores are randomly oriented after deposition. An average acentric orientation and thus a macroscopic EO effect is obtained by a one-time electric-field poling process [36]. To this end, the chip is heated close to the glass transition temperature of JRD1 ($T_g = 82°C$), and a DC poling voltage $U_{pol}$ is then applied across the floating ground electrodes, inducing an electric poling field $E_{pol}$ (green arrows) in the EO material which aligns the dipolar chromophores. This order is frozen by cooling the device to room temperature while maintaining the poling field. The device enables simple push-pull-operation: An RF voltage applied to the GSG transmission line of the MZM induces electric fields in the slots (red arrows) that are parallel to the orientation of the EO chromophores (green arrows) in the right arm and antiparallel in the left arm of the MZM.

The modulation speed of SOH devices may be limited by various aspects such as walk-off between the microwave drive signal and the optical wave, frequency-dependent RF transmission line loss, or RC limitations. Based on previous investigations of very similar devices [37], we expect RC limitations as the dominant effects, originating from the fact that the slot acts as a capacitor which has to be charged and de-charged via the thin doped silicon slabs. To increase the bandwidth of the device, the resistivity of the doped silicon slabs can be decreased by generating an electron accumulation layer [23]. In the experiment, this is realized by applying a gate voltage $U_{gate}$ between the Si substrate and the ground electrode. For the 1.5 mm-long devices used in our experiments, we have measured the electro-optic bandwidth at an externally applied gate field of 0.09 V/nm, leading to a 6 dB bandwidth of approximately 43 GHz. This bandwidth can be improved by further increasing the conductivity of the thin silicon slabs, which can, e.g., be achieved by increasing the doping concentration in these regions. To quantify the modulation efficiency, we measure the static π-voltage of the device, see Supplement 1 for details on the experimental setup. A low-frequency triangular waveform obtained from a function generator (FG) is applied to the signal electrode of the SOH MZM, which is fed by an optical CW signal at a wavelength of 1550 nm through an on-chip grating coupler. The modulated optical signal is detected by a low-speed photo-diode (PD). Both, the output of the FG (blue curve) and the output of the PD (red curve) are recorded simultaneously by an oscilloscope, see Fig. 2 (c) and (d). The modulator is biased in its quadrature point, and the peak-to-peak drive voltage amplitude is chosen larger than the π-voltage such that the MZM is over-modulated. The π-voltage then corresponds to the voltage difference needed to drive the MZM from its minimum to its maximum transmission point.

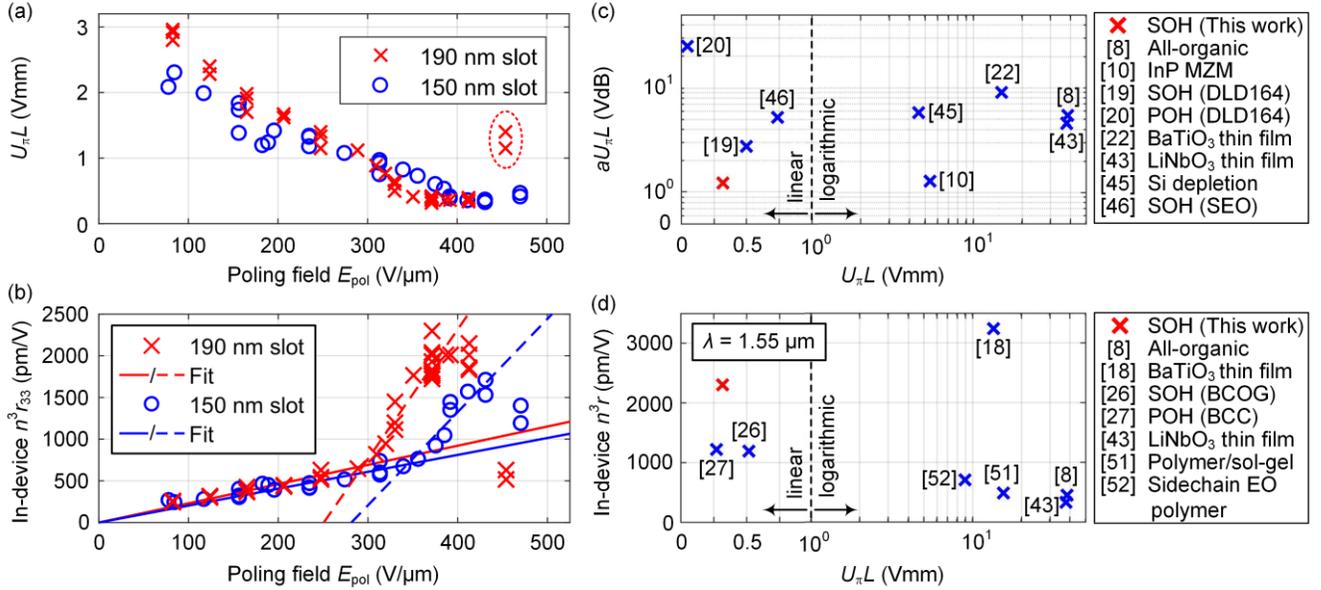

Fig. 3 Electro-optic characterization and benchmarking of the SOH MZM. **(a)** Measured $U_\pi L$-products as a function of the poling field $E_{pol}$. Two sets of devices are investigated: Red crosses correspond to samples with slot widths of 190 nm, and blue circles indicate slot widths of 150 nm. The $U_\pi L$-products decrease for an increasing poling field, level off, and rise again. The rise is attributed to the onset of conductance, possibly in combination with dielectric breakdown, which may cause permanent damage of the affected chromophores. This is confirmed by the fact that the samples indicated by the large red ellipse do not recover typical EO activity when re-poling at moderate fields of 200 V/µm is attempted. The minimum $U_\pi L$ product amounts to 0.32 Vmm for both sample geometries. **(b)** Calculated $n^3 r_{33}$ as a function of $E_{pol}$. Red crosses (blue circles) indicate samples with slot widths of 190 nm (150 nm). For both data sets, $n^3 r_{33}$ increases linearly with the poling fields and features a steep increase in slope once a certain threshold poling field is exceeded. The behavior below (above) the threshold poling field is visualized by the red and blue solid (dashed) lines obtained from least square fits to linear functions. The drop of $n^3 r_{33}$ at even higher poling fields is attributed to the onset of conductance in the OEO material, possibly in combination with dielectric breakdown. The threshold effect may be explained by surface-chromophore interactions leading to alignment of chromophores along the slot sidewalls – this aspect is subject to theoretical investigations. The maximum $n^3 r_{33}$ amounts to 2300 pm/V and is obtained for a slot width of 190 nm. Devices with 150 nm wide slots feature systematically lower values of $n^3 r_{33}$, which is attributed to a stronger influence of surface effects in narrower slots. **(c)** Measured $aU_\pi L$-products as a function of the $U_\pi L$-product for various modulator concepts. The abscissa exhibits a linear scale for $U_\pi L < 1$ Vmm and a logarithmic scale for $U_\pi L \geq 1$ Vmm. Indicated by the proximity to the origin of the figure, SOH MZM combine low voltage-length products with low losses, and hence compare favorably with competing modulator concepts. The device presented in this works shows a $U_\pi L$ of 0.32 Vmm and an $aU_\pi L$ of 1.2 VdB. Only POH modulators offer a lower $U_\pi L$-product of 0.05 Vmm, which comes, however, at the price of an increased $aU_\pi L$-product of 25 VdB **(d)** In-device (EO) figure of merit $n^3 r$ as a function of $U_\pi L$ for competing material platforms for integrated EO modulators. The abscissa exhibits a linear scale for $U_\pi L < 1$ Vmm and a logarithmic scale for $U_\pi L \geq 1$ Vmm. It can be seen that slot waveguide modulators based on organic EO materials offer highest efficiency: Although the in-device EO figures of merit achieved in SOH and POH devices are lower compared to BaTiO$_3$ modulators, the achieved $U_\pi L$ products are almost two orders of magnitude smaller. Note that for all depicted devices except the BaTiO$_3$ modulator, the coefficient $r$ in $n^3 r$ refers to the $r_{33}$ element of the corresponding EO tensor. For the BaTiO$_3$ device an effective EO coefficient which neglects the tensor nature is used since individual tensor components could not be measured. All values given in (d) refer to an operating wavelength of 1550 nm.

For achieving high electro-optic activity, one of the essential parameters is the applied electric poling field $E_{pol}$. In our experiments, we systematically varied this parameter and measured the resulting $U_\pi L$-product for two different sets of devices having slot widths of 190 nm and 150 nm. The experimental results are summarized in Fig. 3(a), where red crosses (blue circles) correspond to slot widths of 190 nm (150 nm). For both sets of devices, the measured $U_\pi L$-products first decrease with increasing $E_{pol}$, then level off, and finally rise again. The rise is attributed to the onset of conductance, possibly in combination with dielectric breakdown. This can inflict permanent damage of the material, which is found by attempting to re-pole the two devices indicated by the red ellipse. In the re-poling experiment, we use moderate poling fields of 200 V/µm, leading to measured $U_\pi L$-products of 3.5 Vmm and 4.0 Vmm. This is much worse than the values of approximately 1.7 Vmm that are typically achieved for this poling field. For both sets of devices, the lowest measured $U_\pi L$-product amounts to 0.32 Vmm, which was achieved at poling fields of 370 V/µm (430 V/µm) for the 190 nm (150 nm) slots. To the best of our knowledge this is the lowest value of $U_\pi L$ hitherto demonstrated for a non-resonant device based on low-loss dielectric waveguides. The propagation loss of the phase-shifter section amounts to about 3.9 dB/mm, see Supplement 1, leading to an $aU_\pi L$ value of 1.2 VdB. This is on par with ultra-low $aU_\pi L$ values of 1.3 VdB achieved for advanced InP-based phase modulators, which combine a rather large $U_\pi L$-product of 5.4 Vmm with a very low on-chip propagation loss of 2.4 dB/cm [10].

Using the simulated fields of the optical and the radio frequency (RF) mode of the device, the measured $U_\pi L$-products in Fig. 3(a) can be linked to the in-device EO figure of merit $n^3 r_{33}$, see Supplement 1 for a derivation of the corresponding mathematical relations. The results are shown in Fig. 3(b), where the calculated $n^3 r_{33}$ is plotted versus the applied poling field. The red crosses and blue circles correspond again to devices having slot widths of 190 nm and 150 nm, respectively. Remarkably, when increasing the poling field, $n^3 r_{33}$ does not exhibit a linear increase with a single slope, as typically observed [38]. Instead, both data sets show a steep increase in slope once a certain threshold poling field is exceeded. To illustrate the two distinctively different poling regimes, we fit a linear function to the measured $n^3 r_{33}$ values below (solid line) and above (dashed line) the threshold poling field. At very high poling fields, we observe a roll-off and finally a drop of $n^3 r_{33}$, which is attributed to the onset of conductance, possibly in combination with dielectric breakdown. These data points are omitted for the linear fit. The unusual dependency of $n^3 r_{33}$ on $E_{pol}$ is still under investigation. It may be explained by chromophore interactions at surfaces - recent simulations have shown that chromophores at the sidewalls of POH slot waveguides align in parallel to these sidewalls and do not contribute to the EO activity [20]. As a consequence, poling is only effective at a sufficient distance from the surface, i.e., in the center of the slot region, where the chromophores are initially randomly oriented and become aligned during poling. Most likely, similar effects are present in SOH slot waveguides. It is conceivable that for sufficiently large poling fields, a threshold is reached that triggers the reorientation of chromophores at the sidewalls, leading to a steeper slope of $n^3 r_{33}$. For a better understanding of the underlying mechanisms, further investigations are required and will be subject to a joint theoretical and experimental effort in the near future. To the best of our knowledge the effect has never been reported so far. We attribute this to the fact that in typical sample geometries the maximum applicable poling field is restricted to 100-200 V/μm before dielectric breakdown occurs [34,39,40]. This is well below the threshold poling field observed in our study. The outstanding stability against dielectric breakdown in SOH devices was already observed in previous experiments where fields above 300 V/μm could be applied without dielectric breakdown [26]. This was attributed to the low number of defects in the EO material filling the nano-scale slot. The maximum calculated $n^3 r_{33}$ is achieved for a sample with slot width of 190 nm and amounts to $(2300 \pm 170)$ pm/V. This result represents not only the highest material-related in-device EO figure of merit for organic EO materials at any operating wavelength but also the highest in-device EO figure of merit ever achieved in a high-speed Pockels-type modulator. Devices with slot widths of 150 nm show systematically smaller values of $n^3 r_{33}$. The largest value achieved amounts to only $(1700 \pm 170)$ pm/V. We attribute this to an even higher fraction of chromophores near the surface than in the center of the slot, which limits the EO activity. The specified error bounds are dominated by the uncertainties of the slot width measurement and the refractive index of JRD1 in the nanoscopic slot waveguide, see Fig. S5 and the corresponding discussion in Supplement 1. In principle, if the refractive index of the EO material is known, the in-device EO coefficient $r_{33}$ can be calculated from the in-device EO figure of merit $n^3 r_{33}$. If we assume the estimated in-device refractive index of JRD1 of $n = 1.81$ we obtain an EO coefficient $r_{33} = 390$ pm/V for the device with 190 nm wide slot, see Supplement 1. However, this value has to be taken with caution, since the refractive index is subject to large uncertainties which are related to birefringence caused by poling-induced chromophore alignment and surface interaction with the sidewalls of the slot waveguide, see Supplement 1.

Note that, for a slot width of 190 nm, the reported value of the in-device $n^3 r_{33}$ is close to the bulk value [28] when taking into account the wavelength dependency of $n^3 r_{33}$. It is hence unlikely that increasing the slot width further will lead to a significant increase of the in-device $n^3 r_{33}$. Additionally, larger slot widths would lead to decreased optical-electrical interaction, see Fig. S4 in Supplement 1, and a smaller RF drive field strength for a given drive voltage. Accordingly, despite the reduced $n^3 r_{33}$, the MZM with 150 nm slots exhibit the same value for $U_\pi L$ as the devices with 190 nm slots, see Fig. 3(a): The devices with narrow slot width exhibit higher drive field strengths as well as a larger overlap of the optical and RF mode, which results in a more efficient modulation and compensates the reduced $n^3 r_{33}$. Note that the threshold effect is less visible in the measurement of the $U_\pi L$-product in Fig. 3(a) due to the inverse relation of $n^3 r_{33}$ and $U_\pi L$.

For a benchmarking of the presented performance parameters of the SOH MZM, we plot the loss-efficiency product $a U_\pi L$ as a function of the voltage-length product $U_\pi L$ for competing state-of-the-art EO modulators, which are based on various concepts and technologies. The results are shown in Fig. 3(c) for an operating wavelength of 1.55 μm, unless stated otherwise. Note that the abscissa features a linear scale for $U_\pi L < 1$ Vmm and a logarithmic scale for $U_\pi L \geq 1$ Vmm. Ideal devices combine low values of $U_\pi L$ with low losses and hence low $a U_\pi L$-products, to be found in the lower left-hand corner of the plot, close to the origin. Besides Pockels-effect modulators, the comparison in Fig. 3(c) also includes selected high-speed InP and silicon-based depletion-type devices. In the future, this group might be complemented by highly efficient accumulation-type devices [41,42], which are currently still subject to limited EO modulation bandwidth of a few gigahertz.

The comparison in Fig. 3(c) shows that modulators fabricated on the thin-film LiNbO$_3$ platform [43] as well as all-organic modulators [8] feature comparatively high $U_\pi L$-products in excess of 38 Vmm, but show also low propagation losses. Hence, despite their large footprint, these devices exhibit moderate $a U_\pi L$-products of around 5 VdB. For the novel BaTiO$_3$-on-Si platform, MZM with lower $U_\pi L$-product of 15 Vmm have been demonstrated [22]. While previous device generations suffered from pronounced propagation losses in excess of 4.5 dB/mm, passive waveguides with losses of only 0.6 dB/mm were demonstrated recently [44], potentially leading to loss-efficiency products of $a U_\pi L = 9$ VdB. State-of-the-art depletion-type Si MZM with U-shaped pn junctions operated at 1310 nm show more efficient modulation with $U_\pi L$-products of 4.6 Vmm and reduced $a U_\pi L$-products down to 5.8 VdB [45]. Similar π-voltage length products of $U_\pi L = 5.4$ Vmm are achieved in indium phosphide (InP) based MZM [10]. At the same time, these devices show significantly reduced propagation losses as compared to silicon-based depletion-type modulators, leading to very low $a U_\pi L$ values of only 1.3 VdB. SOH MZM stand out due to their low $U_\pi L$-products of 0.74 Vmm [46] and 0.5 Vmm [19], obtained by the commercial OEO material SEO100 and by the even more efficient neat chromophore DLD164, respectively. However, the propagation loss of SOH slot waveguides is also higher than the losses of advanced InP waveguides, leading to $a U_\pi L$-products of 5 VdB and 2.8 VdB for the two EO materials. The SOH modulators presented in this work outperform these devices by exhibiting $U_\pi L$-products as low as 0.32 Vmm and a propagation loss of about 3.9 dB/mm, which results in a $a U_\pi L$-product of 1.2 VdB. This is on par with the $a U_\pi L$-product of advanced InP modulators, which feature significantly higher $U_\pi L$-products. The $U_\pi L$-products of POH are even smaller than those of SOH devices, with published values down to 0.05 Vmm [20] when the OEO material DLD164 is employed. However, the small $U_\pi L$-products come at the price of increased propagation losses in the plasmonic slot waveguides, which is reflected in the increased loss-efficiency product of typically $a U_\pi L = 25$ VdB.

Hence, in comparison to all other devices, the SOH modulator presented in this work comes closest to the origin, indicating an ideal combination of high modulation efficiency and low propagation loss of the dielectric waveguide structure. We hence expect that the SOH concept will have major impact on future implementations of optical communication systems and may also open new opportunities in ultra-fast photonic-electronic signal processing, where highly efficient low-loss devices can, e.g., enable advanced electro-optic sampling. SOH devices have previously been demonstrated to enable high-speed coherent [47,48] and non-coherent [24,49] transmission with single-polarization line rates of up to 400 Gbit/s (100 GBd 16QAM) [50].

In Fig. 3(d) we plot the in-device EO figure of merit $n^3 r$ versus $U_\pi L$ for state-of-the-art Pockels-effect EO modulators. For all depicted devices except the BaTiO$_3$ modulator, the EO coefficient $r$ refers to the $r_{33}$ element of the corresponding EO tensor. For the BaTiO$_3$ device, the tensor components could not be measured individually due to the poly-domain structure of the thin film. Instead, $r$ was estimated as an effective EO coefficient by neglecting the tensor nature [18]. Note that the in-device EO figures of merit $n^3 r$ depicted in Fig. 3(d) are purely material-related and do not comprise additional enhancements that can, e.g., be achieved through slow-light structures [51]. For a better comparison between different EO materials, we examine only the values obtained at the standard telecommunication wavelength of 1550 nm. Note that $n^3 r$ is in general wavelength-dependent and usually increases towards smaller wavelengths.

Due to the inverse relation of $U_\pi L$ and $n^3 r$, the lowest value for $U_\pi L$ is expected for the largest $n^3 r$. However, $U_\pi L$ additionally depends on the efficiency of the phase shifter geometry, the limits of which are determined also by the EO material processing technology.

The first class of EO modulators is based on crystalline materials, for which the implementation on well-established photonic platforms such as Si photonics was considered challenging and cumbersome. Recently, however, thin films of LiNbO$_3$ were fabricated and bonded to various substrates such as SOI wafers [17]. In this approach, the EO properties of bulk LiNbO$_3$ can be maintained, leading to devices with EO figures of merit of $n^3 r_{33}$ = 330 pm/V [43] and with moderate π-voltage-length products of $U_\pi L$ = 38 Vmm. The so far highest in-device EO figure of merit of $n^3 r$ = 3200 pm/V is realized on the hybrid BaTiO$_3$-on-SOI platform [18]. Here, the crystalline BaTiO$_3$ is grown by a sophisticated molecular beam epitaxy process, which requires simple phase shifter structures and results in moderate modulation π-voltage-length products of $U_\pi L$ = 13.5 Vmm. Much simpler processing is achieved with OEO materials, which additionally open the possibility of theory-guided molecular design for tailoring the optical properties. All-polymer modulators reach $n^3 r_{33}$ = 450 pm/V [8] when implemented in rather inefficient low index-contrast EO waveguides featuring $U_\pi L$ = 38 Vmm. More efficient modulation with $U_\pi L$ products down to 15.6 Vmm is realized in hybrid polymer/sol-gel modulators [52]. This is on par with BaTiO$_3$ modulators, while the EO figure of merit $n^3 r_{33}$ = 490 pm/V is almost an order of magnitude smaller. This again indicates the advantages of OEO modulator concepts. Modulators based on ultra-thin silicon waveguides clad in thermally stable side-chain EO polymers reach $n^3 r_{33}$ = 710 pm/V [53], which decreases the π-voltage-length product to 9 Vmm. The highest $n^3 r_{33}$ for an SOH modulator so far was achieved with the binary chromophore organic glass (BCOG) PSLD41/YLD124 and amounts to $n^3 r_{33}$ = 1200 pm/V [26], resulting in a $U_\pi L$ product down to 0.52 Vmm. The largest in-device EO figure of merit for a POH modulator amounts to $n^3 r_{33}$ = 1220 pm/V, and is achieved for a binary chromophore composite (BCC) [27]. The π-voltage-length product amounts to $U_\pi L$ = 0.28 Vmm and is significantly higher than the one reported in [20] since a wider slot width of 200 nm was used. Note that the efficiency of the POH phase shifter concept comes at the price of an increased propagation loss, see Fig. 3(c). The SOH modulator presented in this work features an in-device EO figure of merit of $n^3 r_{33}$ = 2300 pm/V and an $U_\pi L$ product of only 0.32 Vmm.

In general, due to the strongly dispersive nature of both the EO coefficient and the refractive index, the EO figure of merit can be further increased when the operating wavelength shifts closer to the absorption maximum of the EO material. As an example, for the POH modulator discussed in [27] the in-device $n^3 r_{33}$ increased by roughly 50 % when the device was operated closer to the absorption maximum of the EO material. Similarly, a significantly higher in-device $n^3 r_{33}$ can be obtained if we operate the presented SOH modulator closer to the absorption maximum of JRD1. For example, for an operation wavelength in the O-band, we expect an in-device EO figure of merit of approximately 4000 pm/V, which even exceeds the EO figure of merit achieved in bulk JRD1 samples at 1310 nm [28].

### Data transmission experiment

To demonstrate the viability of the SOH MZM for high-speed data transmission, we generate on-off-keying (OOK) signals at data rates of 40 Gbit/s. The experimental setup is shown in Fig. 4(a). A pseudo random bit sequence of length $2^{15}$-1 at 40 Gbit/s is obtained from an arbitrary waveform generator (AWG) and applied to the device by a microwave probe without further amplification. The MZM is biased at its quadrature point and terminated by an external 50 Ω impedance in order to reduce reflections of the RF field.

An erbium-doped fiber amplifier (EDFA) after the MZM amplifies the optical signal to a constant power of 8 dBm. A 0.6 nm band pass filter (BP) is used to suppress the amplified spontaneous emission noise of the EDFA. The optical signal is then detected by a high-speed photodiode and a real-time oscilloscope. The oscilloscope features a memory size of 2 Gpts and a sampling rate of 80 GSa/s, which results in a maximum recording time of 25 ms. For a data rate of 40 Gbit/s, this corresponds to a maximum number of $1 \times 10^9$ recorded bits. We are hence limited to measuring a bit error ratio (BER) of BER ≥ $10^{-8}$ [54]. In contrast to that, the $Q$ factors of the recorded eye diagrams indicate BER values below this limit. Therefore we use the $Q$-factor to characterize and compare the signal quality across the various signaling experiments.

The total fiber coupling losses amount to 8.9 dB, and the optical on-chip insertion loss of the MZM adds another 8.2 dB. The on-chip losses are caused by non-ideal MMI-couplers, lossy Si strip waveguides and strip-to-slot converters. The Si slot waveguide losses amount to about 3.9 dB/mm and are mainly attributed to the sidewall roughness of the slot. It can hence be assumed that more advanced lithography processes will allow to further reduce these losses in the future. In fact, it has been already shown that asymmetric strip-loaded Si slot waveguides with propagation losses as low as 0.2 dB/mm can be realized [55].

The device used for the data transmission experiment features a $U_\pi$ of 270 mV. In the following experiments we use 130 μs-long recordings, containing roughly $5 \times 10^6$ symbols. Without any equalization and for a drive voltage of 140 mV$_{pp}$, we obtain a measured $Q$ factor of 5.8, see Fig. 4(b). The energy consumption amounts to 2.5 fJ/bit and is dominated by the power dissipation in the 50 Ω terminating resistor. Operating the device without the terminating resistor, i. e., as a purely capacitive load as in [19], would further reduce the energy consumption. In this case, bandwidth is traded for energy consumption. Note that, since the drive voltage is below the actual $U_\pi$ of the device (270 mV), the extinction ratio of 5 dB (ER = 3.2) as measured from the eye diagram is moderately high, leading to a power penalty of $10 \log_{10}((ER+1)/(ER-1)) = 2.8$ dB These numbers can be improved by increasing the drive voltage, which increases the extinction ratio and the modulation amplitude at the output, but not the modulation loss. The eye diagram without equalization and for a drive voltage of 260 mV$_{pp}$ is depicted in Fig. 4(c). It shows a wider opening with a measured $Q$ factor of 7.0. The energy consumption amounts to 8.5 fJ/bit. The extinction ratio increases to 8.2 dB with a reduced power penalty of only 1.3 dB. This is on par with silicon pn-modulators operating at the same symbol rate, where extinction ratios in the range of (7…10) dB are achieved [56–58]. However, the $U_\pi L$ product for the SOH modulator is at least an order of magnitude smaller than for pn-modulators with a comparable extinction ratio. Note that the static extinction ratio is typically 20 dB to 25 dB higher than the measured one for 40 Gbit/s OOK. This is due to inter-symbol interference caused by the limited modulator bandwidth, and by detected optical noise. The high static extinction ratios in excess of 30 dB are also key to ensure low-chirp operation of push-pull MZM [59]. In general, static extinction ratios of better than 30 dB lead to a chirp parameter $\alpha \leq 0.1$, assuming perfect push-pull operation. This estimation is in good accordance with direct chirp measurements of similar SOH devices [24] and compares favorably with typical pn modulators featuring chirp parameters in the range of 0.8 [60].

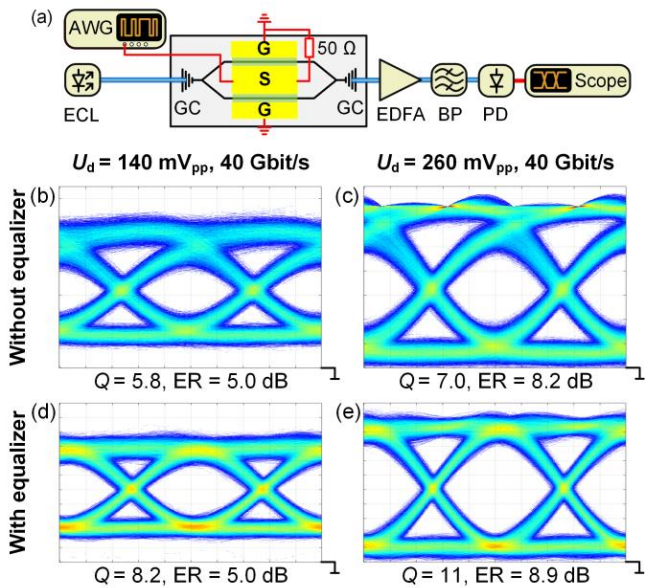

Fig. 4 Data transmission experiments. **(a)** Schematic of the experimental setup. Signals obtained from an arbitrary waveform generator (AWG) are fed to the MZM via microwave probes. The optical carrier provided by an external-cavity laser (ECL) is coupled to and from the chip by grating couplers (GC). The modulated light is amplified by an erbium doped fiber amplifier (EDFA), filtered using a band pass filter (BP), and detected by a high-speed photodiode (PD) connected to a real-time oscilloscope for recording eye diagrams. **(b)** Measured eye diagram at 40 Gbit/s for a drive voltage of 140 mV$_{pp}$ without equalization. The measured $Q$ factor amounts to 5.8 and the extinction ratio (ER) amounts to 5.0 dB. **(c)** Measured eye diagram at 40 Gbit/s for a drive voltage of 260 mV$_{pp}$ without equalization. Both the $Q$ factor and the ER are increased and amount to 7.0 and 8.2 dB, respectively. **(d)** Measured eye diagram at 40 Gbit/s for a drive voltage of 140 mV$_{pp}$ after equalization. From the measured $Q$ factor of 8.2 we estimate a bit error ratio (BER) of $1\times10^{-16}$. **(e)** Measured eye diagram at 40 Gbit/s for a drive voltage of 260 mV$_{pp}$ after equalization. We measure a high $Q$ factor of 11.

The signal quality of the recordings shown in Fig. 4(b) and (c) can be improved by an adaptive filter at the receiver to equalize the detected signals. The filter takes into account the frequency response of the whole signal path including the AWG, the RF cables, the microwave probes, the SOH modulator, the photodiode, and the oscilloscope. The resulting eye diagrams are shown in Fig. 4(d) and Fig. 4(e) for drive voltages of 140 mV$_{pp}$ and 260 mV$_{pp}$, respectively. For the 140 mV$_{pp}$ recording, we obtain a measured $Q$ factor of 8.2. Assuming that after equalization the inter-symbol interference is negligible and that the signal is impaired by additive white Gaussian noise only, this $Q$-factor would lead to a calculated BER of $1 \times 10^{-16}$, which is well below the $10^{-9}$ limit for error-free transmission. To the best of our knowledge, 140 mV$_{pp}$ is the smallest drive voltage ever reported for a system demonstration that achieves comparable $Q$ factors at 40 Gbit/s. For the peak-to-peak drive voltage of 260 mV$_{pp}$ we achieve an even better signal quality with a measured $Q$ factor of 11.

## Summary and Outlook

We demonstrate a silicon-organic hybrid (SOH) Mach-Zehnder modulator (MZM) featuring an ultra-high in-device electro-optic (EO) figure of merit of $n^3r_{33} = 2300$ pm/V by employing the highly efficient EO chromophore JRD1. This is the highest material-related value ever achieved in a high-speed Pockels-type modulator at any operating wavelength. The reported $n^3r_{33}$ is a factor of two larger compared to previously published values for SOH or plasmonic-organic hybrid (POH) modulators operating at the same wavelength of 1550 nm [26,27]. The devices feature ultra-small voltage-length products down to $U_\pi L = 0.32$ Vmm along with loss-efficiency products of $aU_\pi L = 1.2$ VdB. Devices as short as 320 µm may hence be directly driven from binary outputs of energy-efficient CMOS electronics [46], providing voltage swings of the order of 1 V. For these devices, optical propagation losses in the phase shifter sections would amount to less than 1.5 dB, leading to total insertion losses of less than 2 dB when assuming existing implementations of strip-to-slot mode converters [61] and MMI couplers [62] without any further optimization. This compares favorably to competing device concepts such as silicon-based depletion-type pn modulators [63,64] or POH devices [20]. The viability of the presented modulator is demonstrated by generating OOK signals at a data rate of 40 Gbit/s. We obtain high-quality signals with a measured $Q$ factor of 8.2 for an ultra-low peak-to-peak drive voltage of only 140 mV$_{pp}$.

**Funding.** The European Research Council (ERC Starting Grant 'EnTeraPIC', # 280145), the EU FP7 projects PhoxTroT and BigPipes, the Alfried Krupp von Bohlen und Halbach Foundation, the Helmholtz International Research School for Teratronics (HIRST), the Karlsruhe School of Optics and Photonics (KSOP), the Karlsruhe Nano-Micro Facility (KNMF), the National Science Foundation (DMR-1303080) and the Air Force Office of Scientific Research (FA9550-15-1-0319).

**Acknowledgment.** We thank Bruce H. Robinson, Andreas Tillack, and Lewis E. Johnson for helpful discussions. We also thank the anonymous reviewers, whose comments helped in improving the paper.

See Supplement 1 for supporting content.